\documentclass[onecolumn,11pt]{article}
\usepackage[top=1in, bottom=1in, left=1in, right=1in]{geometry}
\usepackage{arxiv}
\usepackage[numbers,sort&compress]{natbib}
\usepackage{setspace}

\title{\vspace{-.2in}Bilinear dynamic mode decomposition\\ for quantum control\vspace{-.1in}}
\author{\normalsize{Andy Goldschmidt$^{1*}$, E. Kaiser$^2$, J. L DuBois$^3$, S. L. Brunton$^2$, and J. N. Kutz$^4$}\\
\footnotesize{$^1$ Department of Physics, University of Washington, Seattle, WA 98195}\\
\footnotesize{$^2$ Department of Mechanical Engineering, University of Washington, Seattle, WA 98195}\\
\footnotesize{$^3$ Lawrence Livermore National Laboratory, Livermore, CA}\\
\footnotesize{$^4$ Department of Applied Mathematics, University of Washington, Seattle, WA 98195}\vspace{-.4in}}
\date{}
\begin{document}
\maketitle
\blfootnote{$^*$ Corresponding author: andygold@uw.edu}

\begin{abstract}
Data-driven methods for establishing {\em quantum optimal control} (QOC) using time-dependent control pulses tailored to specific quantum dynamical systems and desired control objectives are critical for many emerging quantum technologies.  
We develop a data-driven regression procedure, {\em bilinear dynamic mode decomposition} (biDMD), that leverages time-series measurements to establish quantum system identification for QOC.  
The biDMD optimization framework is a physics-informed regression that makes use of the known underlying Hamiltonian structure.  
Further, the biDMD can be modified to model both fast and slow sampling of control signals, the latter by way of stroboscopic sampling strategies.
The biDMD method provides a flexible, interpretable, and adaptive regression framework for real-time, online implementation in quantum systems.  Further, the method has strong theoretical connections to Koopman theory, which approximates nonlinear dynamics with linear operators.  
In comparison with many machine learning paradigms minimal data is needed to construct a biDMD model, and the model is easily updated as new data is collected.  
We demonstrate the efficacy and performance of the approach on a number of representative quantum systems, showing that it also matches experimental results.
\end{abstract}

\section{Introduction} \label{sec:introduction}
{\em Quantum optimal control} (QOC) is a comprehensive mathematical framework for quantum control in which time-dependent control pulses are tailored to specific quantum dynamical systems and desired experimental objectives~\cite{Brif2010}. QOC algorithms are critical for emerging new quantum technologies in scientific and engineering disciplines, including computing, communications, simulation and sensing~\cite{Macfarlane2003}. For example, algorithm construction in the gate-set model of quantum computation is accomplished by concatenation from a universal set of quantum logic gates; however, QOC can create more accurate implementations of commonly used subroutines by leveraging optimal control~\cite{Zahedinejad2015, Wu2020, Shi2020}.
Despite the unique challenges of quantum theory~\cite{Dong2010,Altafini2012}, standard model-based control optimization procedures have been developed and refined~\cite{Brif2010,Khaneja2005,Doria2011,Egger2014}. Many of the seminal innovations in QOC were developed for applications in {\em Nuclear Magnetic Resonance} (NMR)~\cite{Glaser2015}; today there are many numerical methods and tools useful in a wide range of QOC application areas~\cite{Hogben2011,Johansson2013,Ball2020}. 
In QOC, practical control design ultimately depends on an experimentally-accurate model of the governing quantum dynamics and the action of controls. In this manuscript, we introduce the data-driven regression framework known as {\em dynamic mode decomposition} (DMD) to construct control models directly from time-series measurement data. Our DMD method is tailored to the bilinear structure of quantum control dynamics;  moreover, it is a completely data-driven approach that can accommodate both fast and slow (stroboscopic) sampling of control signals.

Many data-driven methods for characterizing quantum devices have been developed under a variety of modelling assumptions~\cite{Eisert2020:Quantum}. The quantum-information community often assumes a quantum-process model of the dynamics. In this case, a data-driven model of a specific quantum process is treated as a {\em black box} and is experimentally characterized using {\em quantum process tomography} (QPT)~\cite{Chuang1997,Mohseni2008,Shabani2011:efficient,Merkel2013,BlumeKohout2017}. An alternative to the QPT black box is {\em Hamiltonian tomography } which attempts to identify the generator of the dynamics by connecting the Hamiltonian parameters to features of the observed dynamics \cite{Geremia2002, Shabani2011:estimation, daSilva2011}. Our approach is similar to Hamiltonian tomography, but we treat control as an exogenous parameter in order to simultaneously identify separate generators for the drift and control dynamics. Therefore, our approach directly addresses the needs of QOC.

The challenges posed by Hamiltonian tomography are well-known~\cite{Eisert2020:Quantum, Shabani2011:estimation}. Statisical difficulties can arise when solving inverse problems involving data. The generated dynamics are a nonlinear function of the Hamiltonian parameters being sought. There is also the {\em curse-of-dimensionality}: computational costs grow exponentially with respect to the quantum degrees of freedom. 
Inspiration can be taken from machine learning methods because they have successfully circumvented many difficulties with high-dimensional and nonlinear data sets through principled construction of low-dimensional embedding spaces. In quantum dynamics, reinforcement learning methods have been used for quantum control design~\cite{Palittapongarnpim2017,Bukov2018,Niu2019}, neural networks have been used for finding low-rank embeddings of quantum states and processes~\cite{Torlai2018,Xin2019,Palmieri2020}, and the efficiencies of Bayesian methods~\cite{Wiebe2014,Wang2017} and system identification algorithms~\cite{Zhang2014} have been studied.

Our approach leverages data-driven system identification: a scalable, accurate, and flexible modeling paradigm ubiquitous throughout science and engineering. System identification algorithms use abundant data collection to automate modeling tasks in frameworks amenable to model-free control design. Koopman theory for classical system identification and control is one paradigm that even uses the operator-theoretic language developed for quantum dynamics~\cite{Koopman1931,Koopman1932,Mezic2005}. In Koopman theory, nonlinear classical dynamics are described by the linear action of Koopman operators on an infinte-dimensional Hilbert space of observables. The action on observables makes the operators amenable to general data analysis. The DMD algorithm is a data-driven regression that makes a finite-dimensional approximation to the Koopman operator~\cite{Tu2014}. DMD avoids the statistical issues associated with inverse problems by finding low-rank, interpretable feature spaces in the data. The algorithm can accommodate physical constraints, delay-embeddings, and multi-scale dynamics~\cite{Askham2018, LeClainche2017, Kutz2016:mrDMD}. With high-dimensional data DMD can make use of tensor decompositions~\cite{Klus2018,Goesmann2020} and compressive sampling~\cite{Brunton2015, Bai2020}. Moreover, the DMD algorithm has been successfully adapted to provide equation-free control design for non-autonomous dynamical systems~\cite{Proctor2016,Macesic2018,Peitz2020,Gosea2020}.  In the present work, we show how the DMD system identification framework can be applied to the study of quantum control via a bilinear DMD algorithm (biDMD)~\cite{Peitz2020, Gosea2020}. We study the biDMD algorithm as an approach to synthesize numerical optimal control and experimentation.  In addition to the innovation of the biDMD algorithm, we also show how it can be used when sampling is fast or slow relative to the applied control signals--the latter by stroboscopic sampling.  The biDMD architecture is applied in several examples in a promising demonstration of the possibility of QOC using time-series data alone.

The first two sections serve as background. Section~\ref{sec:qds} reviews quantum dynamics and recalls the connection with Koopman-von Neumann classical mechanics. Section~\ref{sec:dmd} reviews the DMD algorithm.  In Section~\ref{sec:bidmd}, we introduce biDMD for quantum control. An example is provided for the fast-sample limit where the data record fully resolves the dynamics. Next, Section~\ref{sec:strobe_dmd} explores biDMD in the opposite limit of stroboscopic sampling. In this context, natural interpretations of DMD using Floquet theory and treatments of biDMD from the perspective of average Hamiltonian theory are discussed with examples.

\section{Quantum dynamical systems} \label{sec:qds}
The first complete formulation of quantum dynamical systems occurred in the 1930s with the operator-theoretic perspective of the Dirac-von Neumann axioms~\cite{vonNeumann1932}. This description replaced the notion of a possibly nonlinear equation of motion defined over a classical phase space with an infinite dimensional linear operator algebra acting on a Hilbert space. This formulation was not restricted to quantum mechanics. Indeed, an operator-theoretic linearization of classical dynamical systems was contemporaneously made using the Koopman-von Neumann theory~\cite{Koopman1931,Koopman1932}. We review the dynamics of quantum states here before transitioning to abstract states of measurement data in Section~\ref{sec:dmd}.

\subsection{Autonomous systems} \label{sec:autonomous}
In quantum mechanics, the state of an $N$-dimensional physical system is represented by a unit vector, or ket, $\ket{\psi}$ in a complex vector space $\C^N$. The dynamics of a quantum state are given by the Schr\"{o}dinger equation,
\begin{equation}
    \frac{\partial}{\partial t} \ket{\psi(t)} = -i H_0 \ket{\psi(t)},\qquad \ket{\psi(t_0)} = \ket{\psi_0} \,,
\end{equation}
where the Hamiltonian operator, $H_0$, has been assumed without loss of generality to be traceless , and Plank's constant $\hbar$ has been set to unity. More generally, an ensemble of pure quantum states can be completely characterized, in the sense of its measurement statistics, by a density matrix $\rho(t)$; that is, a non-negative self-adjoint operator in $\C^{N \times N}$ with trace one. The Liouville-von Neumann, or quantum Liouville equation, describes the time evolution of a density matrix~\cite{Breuer2010}:
\begin{equation} \label{eqn:quantumliouville}
    \frac{\partial}{\partial t} \rho(t) = -i [H_0, \rho(t) ] := \mathcal{L}_0 \rho(t),\qquad \rho(t_0) = \rho_0 \,.
\end{equation}
The quantum Liouville operator $\mathcal{L}_0$ is sometimes known as a \textit{super-operator} because of its linear action on the space of operators. There are a number of ways to \textit{vectorize} the quantum Liouville equation. We will use the language of the familiar Bloch vector (also known as the vector of coherence) to describe passing to a vector of differential equations. Writing $\rho(t) = \1/N + \sum_{j=1}^{N^2-1} \Tr(\rho \sigma_j) \sigma_j$ with $\{\sigma_j\}_{j=1}^{N^2-1}$ as a complete and orthonormal basis for traceless Hermitian operators, we can then define $x_j := \Tr(\rho \sigma_j)$ so that (\ref{eqn:quantumliouville}) becomes~\cite{Altafini2012,Kurniawan2012}
\begin{equation} \label{eqn:vecliouville1}
    \frac{\partial}{\partial t} \mathbf{x}(t) = \mathbf{L}_0 \mathbf{x}(t), \qquad \mathbf{x}(t_0) = \mathbf{x}_0 \,.
\end{equation}
For a more complete description of this process, see Appendix~\ref{apdx:vectorize}. Obtaining an expectation value $x_j(t) := \Tr(\rho(t) \sigma_j)$ requires a collection of identically-prepared quantum states. In this paper, the ``measurement'' of a quantum state as represented by $\mathbf{x}(t)$ is understood to mean the measurement of such an ensemble. 

In terms of the eigenvectors $\mathbf{v}_j$ and eigenvalues $\lambda_j$ of $\mathbf{L}_0$, the solution to (\ref{eqn:vecliouville1}) is
\begin{equation} \label{eqn:eigcontinuous}
    \mathbf{x}(t) = \sum_{j=1}^{N^2-1} \mathbf{v}_j e^{\lambda_j (t-t_0)} c_j = \mathbf{V} \exp(\boldsymbol{\Lambda} (t-t_0)) \mathbf{c} \,,
\end{equation}
where the coefficients $c_j$ are the coordinates of $\mathbf{x}_0$ in the eigenvector basis. For continuous dynamics as in (\ref{eqn:eigcontinuous}), there exists an equivalent discrete time description of the system $\mathbf{x}(t)$ sampled at intervals of $\Delta t$.

\subsection{Non-autonomous systems} \label{sec:nonautonomous}
Many important classical non-autonomous control systems can be formulated as control-affine dynamical systems. Transforming to the operator theoretic perspective of the Koopman-von Neumann theory, these are bilinear dynamical systems~\cite{Goswami2017}. In this section, we review non-autonomous dynamical systems and establish the bilinear formulation of the quantum control problem.

\subsubsection{Direct actuation} \label{sec:aut_direct}
Direct actuation is a common situation in classical control theory in which an autonomous dynamical system like (\ref{eqn:vecliouville1}) undergoes a linear response to control inputs $\mathbf{u}(t) \in \R^{N_c}$~\cite{Proctor2016}:
\begin{equation} \label{eqn:vecliouville2}
    \frac{\partial}{\partial t} \mathbf{x}(t) = \mathbf{L}_0 \mathbf{x}(t) + \mathbf{L}_\mathrm{B} \mathbf{u}(t), \qquad \mathbf{x}(t_0) = \mathbf{x}_0 \,.
\end{equation}
Recall that if the control is input under a zero-order hold across $\Delta t$, the discretization $\mathbf{x}_n = \mathbf{x}(t_0 + (n-1) \Delta t)$ and $\mathbf{u}_n = \mathbf{u}(t_0 + (n-1) \Delta t)$ transforms (\ref{eqn:vecliouville2}) into the discrete-time dynamical system
\begin{equation} \label{eqn:directactuation2}
     \mathbf{x}_{n+1} = e^{\mathbf{L}_0 \Delta t} \mathbf{x}_n + \left( \int_0^{\Delta t} e^{\mathbf{L}_0(\Delta t - s)} \mathbf{L}_\mathrm{B} \D s \right) \mathbf{u}_n \,.
\end{equation} 
Significantly, the dynamics remain control-affine for any span $\Delta t$. 

\subsubsection{Bilinear quantum dynamics} \label{sec:aut_bilinear}
The control of a quantum system can be modelled using $N_c$ real-valued control functions,~$u_j(t)$, coupled to corresponding time-independent interaction Hamiltonians,~$H_j$, such that the dynamics are described by the bilinear Schr\"{o}dinger equation,
\begin{equation} \label{eqn:bilinearschroedinger}
    \frac{\partial}{\partial t} \ket{\psi(t)} = -i \left(H_0 + \sum_{j=1}^{N_c} u_j(t) H_j \right) \ket{\psi(t)}, \qquad \ket{\psi(t_0)} = \ket{\psi_0} \,.
\end{equation}
Following Appendix~\ref{apdx:vectorize} like for (\ref{eqn:vecliouville1}), the bilinear Schr\"{o}dinger equation induces the vectorized bilinear quantum Liouville equation~\cite{Altafini2012,Kurniawan2012},
\begin{equation} \label{eqn:bilinearvecliouville}
    \frac{\partial}{\partial t} \mathbf{x}(t) = \left(\mathbf{L}_0 + \sum_{j=1}^{N_c} u_j(t) \mathbf{L}_j \right) \mathbf{x}(t), \qquad \mathbf{x}(t_0) = \mathbf{x}_0 \,,
\end{equation}
by identifying $H_n \mapsto \mathbf{L}_n$ for $n=0,1,\dots,N_c$ and $\mathbf{x} \in \R^{N^2-1}$. This linear differential equation can be integrated in the usual way to obtain
\begin{equation}
    \mathbf{x}(t) = e^{\mathbf{L}_0 (t-t_0)}\mathbf{x}_0 + \sum_{j=1}^{N_c} \int_{t_0}^t e^{\mathbf{L}_0((t-t_0)-s)} u_j(s) \mathbf{L}_j \mathbf{x}(s) \D s .
\end{equation}
If a discretization is taken such that $\mathbf{x}_n = \mathbf{x}(t_0 + (n-1) \Delta t)$ and $u_{j;n} = u_j(t_0 + (n-1) \Delta t)$, then to first order
\begin{equation} \label{eqn:bilineardst}
     \mathbf{x}_{n+1} = (\1 + \mathbf{L}_0 \Delta t) \mathbf{x}_n + \sum_{j=1}^{N_c} u_{j;n} \mathbf{L}_j \Delta t \, \mathbf{x}_n + \mathrm{O}(\Delta t^2) \,,
\end{equation}
such that the discrete time dynamical system remains control-affine~\cite{Peitz2020}. Note that in this first-order approximation the computation of derivatives via finite differences and the computation of the discrete time dynamical system are equivalent.

\section{Dynamic Mode Decomposition} \label{sec:dmd}
%
\begin{figure}[t]
    \includegraphics[width=\textwidth]{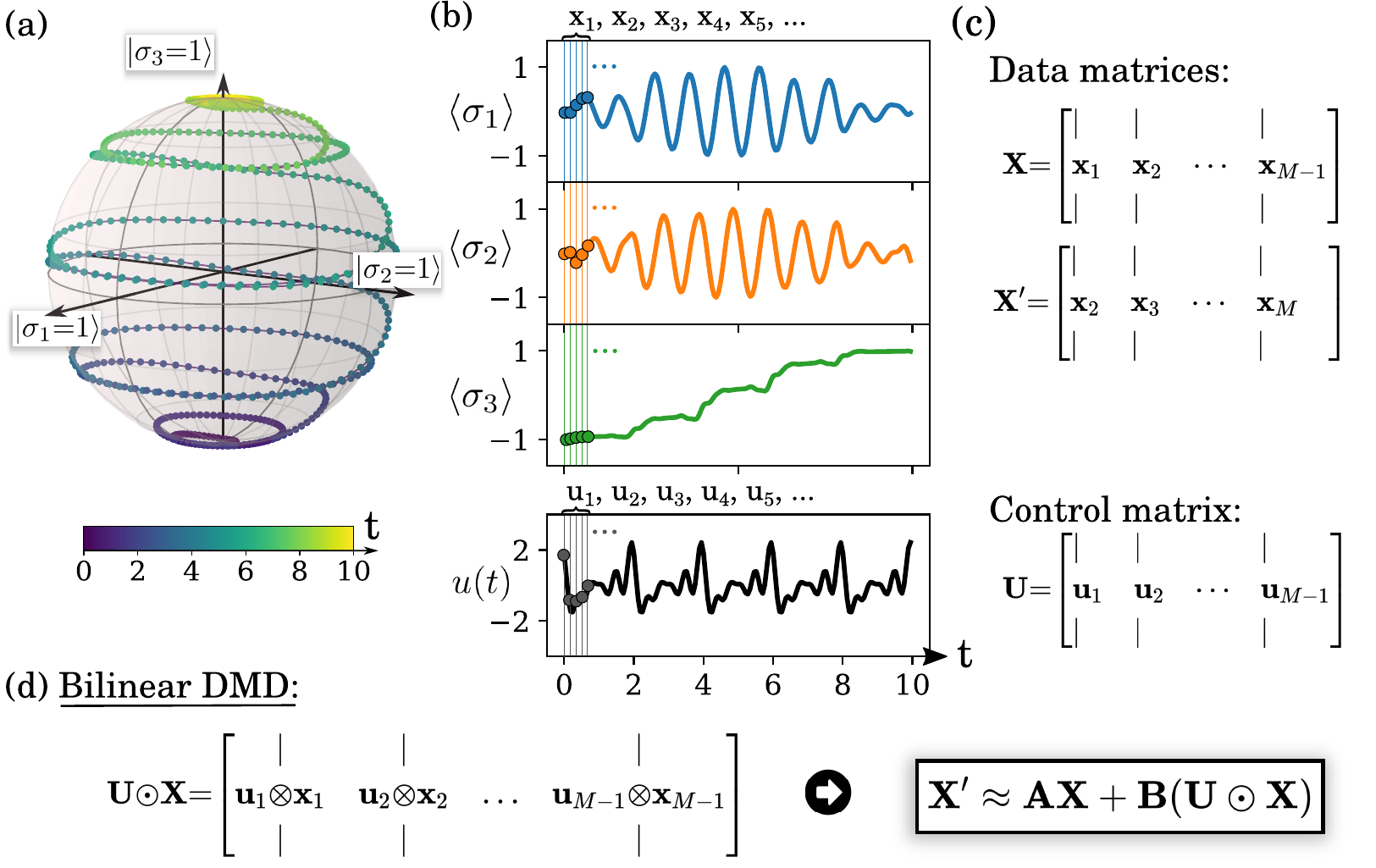}
    \centering
    \caption{
         The trajectory of a qubit driven by a linearly-polarized semi-classical drive $\textrm{u}(t)$ (Hamiltonian: $H(t) = \pi \sigma_z + \textrm{u}(t) \sigma_x$) is shown on the Bloch sphere in Figure~\ref{fig:main1}(a). The corresponding Pauli-spin measurements are shown in Figure~\ref{fig:main1}(b). Measurements $\mathbf{x}_j$, $j=1,2,\dots$, are taken at discrete time steps and assembled into offset snapshot matrices $\mathbf{X}$ and $\mathbf{X}'$ in Figure~\ref{fig:main1}(c). The bilinear Dynamic Mode Decomposition (biDMD, Figure~\ref{fig:main1}(d)) is a regression-based algorithm that uses the assembled data matrices and control input from sufficiently-resolved data to learn the intrinsic dynamics,~$\mathbf{A}$ and the control,~$\mathbf{B}$, for the bilinear dynamics.
    }
    \label{fig:main1}
\end{figure}

In recent years, a variety of practical computational tools have encouraged more widespread adoption of the Koopman-von Neumann perspective in the dynamical systems community~\cite{Mezic2005,Mauroy2020}. The Dynamic Mode Decomposition (DMD) broadly refers to a suite of numerical methods originating in the fluid dynamics community for the purpose of studying coherent spatiotemporal structures in complex fluid flows~\cite{Tu2014}. DMD combines standard dimensionality reduction via the singular value decomposition with Fourier transforms in time. In application, all variations of the DMD algorithm involve collecting a time series of experimental measurements in order to compute a reduced-order dynamical model based on DMD modes and eigenvalues.  The DMD modes identify coherent structures in the measured space and the DMD eigenvalues define the growth, decay, and oscillation frequency of each mode. The following sections will review the DMD framework.

\subsection{DMD} \label{sec:classicdmd}
The standard DMD is a data-driven algorithm defined via the observed trajectories from either experimental data or a numerical simulation of a dynamical system
\begin{equation} \label{eqn:dmdsystem}
    \frac{\partial}{\partial t} \mathbf{x}(t) = \mathbf{L}_0 \mathbf{x}(t), \qquad \mathbf{x}(t_0) = \mathbf{x}_0 \,,
\end{equation}
where $\mathbf{x}(t) \in \R^{N^2-1}$ to remain consistent with Section~\ref{sec:qds}. (DMD ultimately deals with data and is agnostic to the underling system that produced it.) The first step of the algorithm is to assemble a sequential measurement record $\{\mathbf{x}(t_1) = \mathbf{x}_1, \mathbf{x}_2, \dots, \mathbf{x}_{M}\}$, $t_1 < t_2 < \dots < t_m$, into \textit{snapshot matrices}:
\begin{eqnarray} \label{eqn:snapshots}
    &\mathbf{X} :=
    \begin{bmatrix}
        | & | & & | \\
        \mathbf{x}_{1} & \mathbf{x}_{2} & \dots & \mathbf{x}_{M-1} \\
        | & | & & |
    \end{bmatrix}, \qquad
    &\mathbf{X}' :=
    \begin{bmatrix}
        | & | & & | \\
        \mathbf{x}_{2} & \mathbf{x}_{3} & \dots & \mathbf{x}_{M} \\
        | & | & & |
    \end{bmatrix}\,,
\end{eqnarray}
so $\mathbf{X}, \mathbf{X'} \in \R^{N^2-1 \times M-1}$. In this paper we assume uniform sampling with $t_m := t_0 + (m-1) \Delta t$, but this assumption can be relaxed~\cite{Askham2018}. In its simplest form, DMD is a regression algorithm that estimates the propagator matrix
\begin{equation} \label{eqn:dmd}
    \mathbf{X}' \approx \mathbf{A} \mathbf{X}
\end{equation}
by solving the optimization problem 
\begin{equation} \label{eqn:lstsqrs}
    \mathbf{A} = \arg\min_{\hat{\mathbf{A}}} \norm{ \hat{\mathbf{A}} \mathbf{X} - \mathbf{X}'}_F
\end{equation}
where $\norm{\mathbf{M}}_F = \sqrt{\sum_{j=1}^{J}\sum_{k=1}^{K} m_{jk}}$ is the Frobenius matrix norm defined for any given matrix $\mathbf{M}$. The least-squares solution to (\ref{eqn:lstsqrs}) is $\mathbf{A} = \mathbf{X}'\mathbf{X}^+$ where $+$ denotes the Moore-Penrose pseudoinverse of a matrix. From the eigendecomposition $\mathbf{A} = \mathbf{W} \boldsymbol{\Omega} \mathbf{W}^{-1}$, define the DMD modes $\mathbf{w}_j$ as the columns of $\mathbf{W}$. Future states can then be predicted as in a discrete analogue to (\ref{eqn:eigcontinuous}) using 
\begin{equation} \label{eqn:dmdpredict}
    \mathbf{x}_n = \sum_{j=1}^{N^2-1} \mathbf{w}_j \omega^n_j b_j = \mathbf{W} \boldsymbol{\Omega}^n \mathbf{b}
\end{equation}
where $\mathbf{b}$ are the coefficients of the initial condition in the eigenvector basis, $\mathbf{x}_0 = \mathbf{W} \mathbf{b}$.

\subsubsection{Scalability}

The eigendecomposition of $\mathbf{A} \in \R^{N^2-1 \times N^2-1}$ can easily become numerically intractable for large $N$. However, the big data limit is nothing unusual for the problems appearing in data science and fluid dynamics that DMD has already addressed with success. The DMD algorithm circumvents this problem by projecting the high-dimensional snapshot matrix $\mathbf{X}$ onto a low-rank subspace defined by $R$ principle orthogonal decomposition (POD) modes computed from the singular value decomposition of $\mathbf{X}$ (note that because it is computed from the snapshot matrices, the rank of $\mathbf{A}$ is at most $M-1$ when $N^2 > M$). In this way, DMD determines a low-rank approximation $\tilde{\mathbf{A}} \in \R^{R \times R}$ to describe the dynamics of the measured trajectories. This paper introduces DMD for quantum control by way of simple low-rank examples. However, suitable amplitude constraints on the control can ensure the POD-based DMD is amenable to the study of many-body quantum states~\cite{Kosut2020}. As well, DMD accommodates the use of tensor decompositions~\cite{Klus2018,Goesmann2020} and compressive sampling~\cite{Brunton2015}.

\subsection{DMD with Control (DMDc)} \label{sec:dmdc}
Dynamic Mode Decomposition with control (DMDc)~\cite{Proctor2016} is an algorithm developed for modelling the special case of direct actuation:
\begin{equation} \label{eqn:dmdc_system}
    \frac{\partial}{\partial t} \mathbf{x}(t) = \mathbf{L}_0 \mathbf{x}(t) + \mathbf{L}_\mathrm{B} \mathbf{u}(t), \qquad \mathbf{x}(t_0) = \mathbf{x}_0 \,,
\end{equation}
with $\mathbf{u}(t) \in \R^{N_c}$. Like DMD in (\ref{eqn:snapshots}), the size-$M$ measurement record associated with the trajectory of $\mathbf{x}$ is assembled into snapshot matrices $\mathbf{X}$ and $\mathbf{X}'$. This time, the control inputs, $\mathbf{u}_n := \mathbf{u}(t_n)$, are also collected:
\begin{equation} \label{eqn:snapshotu}
    \boldsymbol{U} = \begin{bmatrix}
        | & | & & | \\
        \mathbf{u}_{1} & \mathbf{u}_{2} & \dots & \mathbf{u}_{M-1} \\
        | & | & & |
    \end{bmatrix}\,.
\end{equation}
DMDc is a regression-based approach to system identification that disambiguates the intrinsic dynamics,~$\mathbf{A}$, and the effects of control,~$\mathbf{B}$:
\begin{equation} \label{eqn:dmdc}
   \mathbf{X}' \approx \mathbf{A} \mathbf{X} + \mathbf{B} \mathbf{U}
   = \begin{bmatrix}
        \mathbf{A} & \mathbf{B}
    \end{bmatrix}
    \begin{bmatrix}
        \mathbf{X} \\
        \mathbf{U} \\
    \end{bmatrix} := \mathbf{G} \boldsymbol{\Xi} \,.
\end{equation}
The DMDc algorithm achieves this disambiguation by interpreting the best-fit solution according to
\begin{equation}
    \begin{bmatrix}
        \mathbf{A} & \mathbf{B}
    \end{bmatrix} = \mathbf{X}'\boldsymbol{\Xi}^+ \,.
\end{equation}
DMDc brings all the benefits of the DMD algorithm to a model-free framework for experimental control optimization.

\begin{algorithm}[t]
\caption{Bilinear Dynamic Mode Decomposition}
\label{alg:bidmd}
\begin{algorithmic}[1]
		\Statex{\textbf{INPUT: } Snapshot data $\mathbf{X}$, shifted snapshot data $\mathbf{X}'$, control matrix $\mathbf{U}$, and target ranks $\tilde{r}, \hat{r}$}
		\Statex{\textbf{OUTPUT: } DMD modes $\boldsymbol{\Phi}$, eigenvalues $\boldsymbol{\Lambda}$, and estimates $\widetilde{\mathbf{A}}$, $\widetilde{\mathbf{B}}$ for the drift and control matrices}
		\Function{BIDMD}{$\mathbf{X}$, $\mathbf{X}'$, $\mathbf{U}$, $\tilde{r}$, $\hat{r}$}
		    \State{$\boldsymbol{\Xi} \leftarrow \begin{bmatrix} \mathbf{X} \\ \mathbf{U} \odot \mathbf{X} \end{bmatrix}$}
    		\State{$\widetilde{\mathbf{U}}, \widetilde{\boldsymbol{\Sigma}}, \widetilde{\mathbf{V}} \leftarrow \text{SVD}(\boldsymbol{\Xi}, \tilde{r})$}
    		    \Comment{Truncated $\tilde{r}$-rank SVD of $\boldsymbol{\Xi}$}
    		\State{$[\widetilde{\mathbf{U}}_A, \widetilde{\mathbf{U}}_B ] \leftarrow \widetilde{\mathbf{U}}$ }
    		    \Comment{Decompose $\widetilde{\mathbf{U}}$ according to $\mathbf{A}$ and $\mathbf{B}$}
            \State{$\widetilde{\mathbf{A}} \leftarrow  \mathbf{X}' \widetilde{\mathbf{V}} \widetilde{\boldsymbol{\Sigma}}^{-1} \widetilde{\mathbf{U}}_A^\dagger $}
                \Comment{Estimate for $\mathbf{A}$}
            \State{$\widetilde{\mathbf{B}} \leftarrow  \mathbf{X}' \widetilde{\mathbf{V}} \widetilde{\boldsymbol{\Sigma}}^{-1} \widetilde{\mathbf{U}}_B^\dagger $}
                \Comment{Estimate for $\mathbf{B}$}
    		\State{$\widehat{\mathbf{U}}, \widehat{\boldsymbol{\Sigma}}, \widehat{\mathbf{V}} \leftarrow \text{SVD}(\mathbf{X}', \hat{r})$}
                \Comment{Truncated $\hat{r}$-rank SVD of $\mathbf{X}'$}
            \State{$\widehat{\mathbf{A}} \leftarrow \widehat{\mathbf{U}}^\dagger \widetilde{\mathbf{A}} \widehat{\mathbf{U}}$} 
                \Comment{Low-rank approximation for $\mathbf{A}$}
    		\State{$\mathbf{W}, \boldsymbol{\Lambda} \leftarrow \textrm{EIG}(\widehat{\mathbf{A}})$ }
    		    \Comment{Eigendecomposition of $\widehat{\mathbf{A}}$}
    		\State{$\boldsymbol{\Phi} \leftarrow \widetilde{\mathbf{A}} \widehat{\mathbf{U}} \mathbf{W}$}
    		    \Comment{DMD modes for $\mathbf{A}$}
    	\EndFunction
\end{algorithmic}
\end{algorithm}
\subsection{Bilinear DMD (biDMD)} \label{sec:bidmd}

Bilinear Dynamic Mode Decomposition (biDMD, Algorithm~\ref{alg:bidmd}) is a data-driven system identification framework for bilinear non-autonomous control systems:
\begin{equation} \label{eqn:bidmd_system}
    \frac{\partial}{\partial t} \mathbf{x}(t) = \left(\mathbf{L}_0 + \sum_{j=1}^{N_c} u_j(t) \mathbf{L}_j \right) \mathbf{x}(t) \,,  \quad \mathbf{x}(t_0) = \mathbf{x}_0 \,.
\end{equation}
Construct the snapshot matrices $\mathbf{X}$, $\mathbf{X}'$, and $\mathbf{U}$ as in (\ref{eqn:snapshots}) and (\ref{eqn:snapshotu}) using the size-$M$ measurement record of the trajectory $\mathbf{x}(t)$ and the corresponding input record $\mathbf{u}(t)$. Additionally, construct the bilinear snapshot matrix
\begin{equation} \label{eqn:snapshotbi}
    \mathbf{U}\odot\mathbf{X} = \begin{bmatrix}
        | & | & & | \\
         \mathbf{u}_{1} \otimes \mathbf{x}_{1} & \mathbf{u}_{2} \otimes \mathbf{x}_{2} & \dots & \mathbf{u}_{M-1} \otimes \mathbf{x}_{M-1} \\
        | & | & & |
    \end{bmatrix} \,.
\end{equation}
where $ \mathbf{u} \otimes \mathbf{x} := [u_{1} x_{1}, u_{1} x_{2}, \dots, u_{1} x_{N^2-1}, u_{2} x_{1}, u_{2} x_{2}, \dots, u_{N_c} x_{N^2-1}]\T$ is the Kronecker product and $\mathbf{U}\odot\mathbf{X}$ is the column-wise Kronecker product of two matrices (also known as the Khatri–Rao product). The discrete time dynamics of a bilinear dynamical system are well-approximated (for small enough $\Delta t$)~\cite{Peitz2020} by
\begin{equation} \label{eqn:bidmd1}
    \mathbf{X}' \approx \mathbf{A} \mathbf{X} + \mathbf{B} \left( \mathbf{U}\odot\mathbf{X} \right) \,.
\end{equation}
where $\mathbf{X} \in \R^{N^2-1 \cross M-1}$, $\mathbf{U} \in \R^{N_c \cross M-1}$, $\mathbf{U}\odot\mathbf{X} \in \R^{N_c (N^2-1) \times M-1}$, $\mathbf{A} \in \R^{N^2-1 \times N^2-1}$, and $\mathbf{B} \in \R^{N^2-1 \times N_c(N^2-1)}$. In its simplest form, the biDMD algorithm is a regression algorithm in the spirit of DMD (\ref{eqn:dmd}) and DMDc (\ref{eqn:dmdc}) such that
\begin{equation} \label{eqn:bidmd2}
    \mathbf{X}' \approx 
    \begin{bmatrix}
        \mathbf{A} & \mathbf{B}
    \end{bmatrix}
    \begin{bmatrix}
        \mathbf{X} \\
        \mathbf{U}  \odot \mathbf{X} \\
    \end{bmatrix} := \mathbf{G} \boldsymbol{\Xi}
\end{equation}
Like DMDc, the biDMD algorithm (Algorithm~\ref{alg:bidmd}) disambiguates the effect of the intrinsic dynamics,~$\mathbf{A}$, from the effect of the control inputs,~$\mathbf{B}$, using a factorization of the best-fit solution:
\begin{equation} \label{eqn:bidmd3}
    \begin{bmatrix}
        \mathbf{A} & \mathbf{B}
    \end{bmatrix} = \mathbf{X}'\boldsymbol{\Xi}^+ \,.
\end{equation}

\subsubsection{Example 1} \label{sec:example1}
%
\begin{figure}
    \includegraphics[width=\textwidth]{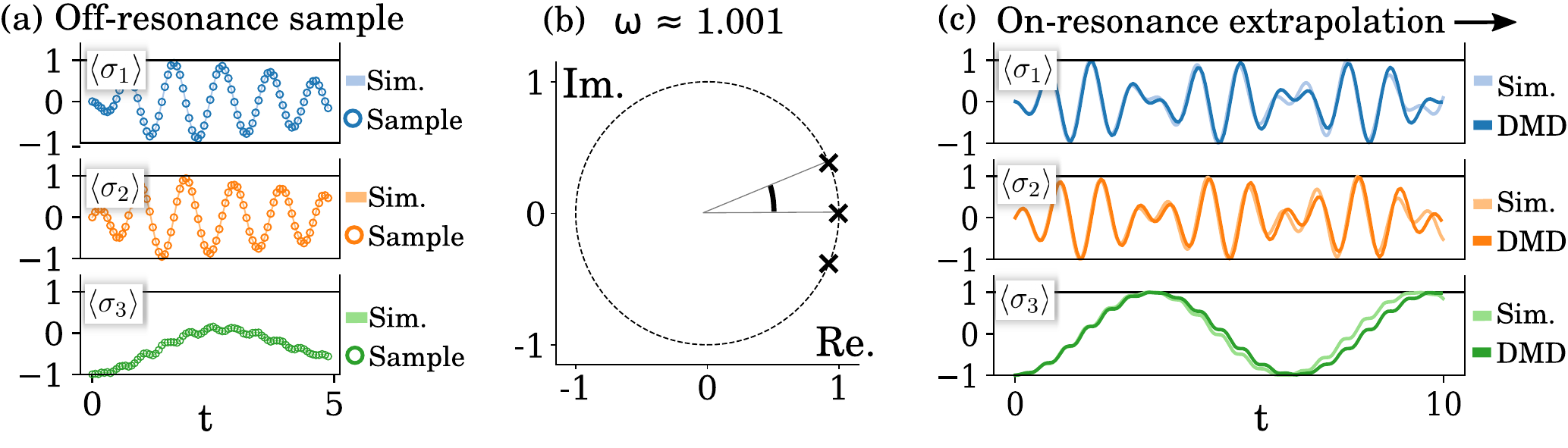}
    \caption{
        In Figure~\ref{fig:bidmd}(a) five periods ($T=1$) of highly-resolved Pauli-spin data ($\Delta t = T/16$) with additive Gaussian noise ($\sigma = 0.01)$ are sampled from simulations of the system $H(t) = \pi \sigma_z + \cos(2\pi\omega_D) \sigma_x$ driven by the slightly off-resonance frequency $\omega_D = 1.1$.  The biDMD algorithm disambiguates the drift, $\mathbf{A}$, and control, $\mathbf{B}$, dynamical operators. The biDMD eigenvalues of $\mathbf{A}$ are plotted in Figure~\ref{fig:bidmd}(b). In Figure~\ref{fig:bidmd}(c) a new control is provided using the resonance frequency, $\omega \approx 1.001$, estimated from the DMD eigenvalues in Figure~\ref{fig:bidmd}(b). The biDMD model extrapolates using only an initial state and a desired control. Figure~\ref{fig:bidmd}(c) compares the biDMD extrapolation to the known simulation.
        }
    \label{fig:bidmd}
\end{figure}
Consider a two-level quantum system, $H(t) = \pi \sigma_3 + u(t) \sigma_1$, where $\{\sigma_j\}_{j=1}^{3}$ are the standard Pauli matrices. This can be realized, to take an example from quantum optics, by applying a linearly polarized semi-classical drive to a dipole-allowed atomic transition $\Delta E = 2 \pi$~\cite{Breuer2010}. Suppose the dynamics have been inaccurately characterized so that the system is driven in an experiment by a slightly off-resonant pulse, $u(t) = \cos(2 \pi \omega_D t)$ with $\omega_D = 1.1$. In Figure~\ref{fig:bidmd}, the biDMD algorithm uses samples from such an experiment in order to disambiguate the effect of control. The phase of the DMD eigenvalues (\ref{eqn:dmdpredict}) provide an estimate of the resonance frequency, $\omega \approx 1.001$. The biDMD model can then be used to predict the behavior of the system under the influence of this estimated-resonance drive without a~priori obtaining an accurate characterization.

Recall the trajectory samples are measured via the expected value of an ensemble of identically prepared quantum states. Advantageously, the DMD algorithm is based on a least-squares regression that is optimal with respect to Gaussian measurement noise. The algorithm does not require much data. Moreover, if the algorithm is applied to a system with dissipation, then the fit DMD eigenvalues are simply forced into the unit circle according to the modified drift operator. An otherwise identical result to Figure~\ref{fig:bidmd} is obtained without any modification to the framework.

\section{DMD for Stroboscopic Sampling} \label{sec:strobe_dmd}
Recall that the DMD algorithms in Section~\ref{sec:dmd} are applicable only to systems in which the measurement resolution $\Delta t$ is small enough to accommodate the linear approximation to the desired accuracy. Higher-order approximations to (\ref{eqn:bilineardst}) show that improvements to the biDMD model can be made by adding terms nonlinear in the control~\cite{Peitz2020}. This research direction may be appropriate for cases where the zero-order hold on the control is extended for the entirety of the interval $\Delta t$. Instead, in this section we study the case of stroboscopic measurements of the trajectory $\mathbf{x}(t)$ separated by time steps $T \gg \Delta t$ during which the control can change significantly.

The observation of low-frequency dynamics $\sim T$ in a system under the influence of relatively high-frequency actuation $\sim \Delta t$ is a familiar feature of bilinear systems. For example, the classroom experiment of an inverted pendulum stabilized by the high-frequency vertical drive of its pivot can be modelled as a bilinear control system (under an appropriate operator-theoretic transformation of Mathieu's equation). The two-level quantum system from Section~\ref{sec:example1} provides another example. The textbook analysis of the two-level system proceeds via a rotating frame transformation together with the rotating wave approximation (RWA). Consider $H(t) = \pi \sigma_3 + u(t) \sigma_1$ driven by a pure tone $u(t) = u_0 \cos(2 \pi \nu t)$. Changing to a frame rotating with the drive frequency about the $\sigma_3$ axis of the Bloch representation,
\begin{equation} \label{eqn:rwa}
    \tilde{H} = U^\dagger H U - i U^\dagger \frac{\partial U}{\partial t} 
    = \underbrace{\pi \nu \1 + \pi(1-\nu) \sigma_3 + \frac{u_0}{2} \sigma_1}_\textrm{RWA}  + \frac{u_0}{2}(\cos(4 \pi \nu t) \sigma_1 -  \cos(4 \pi \nu t) \sigma_2)
\end{equation}
with $U(t) = \exp(2 \pi \nu i (\1 - \sigma_3)t/2)$ where the RWA disregards the contribution of the emergent high-frequency terms to realize a constant Hamiltonian. The coefficient of $\sigma_x$ in the RWA is $u_0$ which can be small relative to $2\pi\nu$. In this case, the characteristic Rabi cycle of the two-level system is a low-frequency oscillation divorced from the high-frequency drive. Motivated by these examples, Sections~\ref{sec:floquet}-\ref{sec:ffm} discuss DMD strategies for the case of stroboscopic measurements.

\subsection{DMD and Floquet Theory} \label{sec:floquet}
In this section, we recall how periodically-driven dynamical systems are connected to the case of stroboscopic measurements (and rotating frames) using the framework of Floquet theory. We introduce Floquet theory and show how corresponding re-interpretations of DMD enable an efficient method for studying stroboscopically-measured, periodically-driven quantum systems.

In Floquet theory (known as Bloch theory in condensed matter physics), the discrete-time evolution of a periodic quantum dynamical system, $H(t+T) = H(t)$, is given by the time-independent stroboscopic Floquet Hamiltonian~\cite{Bukov2015},
\begin{equation} \label{eqn:floquet}
    U(t_0 + T, t_0) = \exp(- i H_F[t_0] T) \,.
\end{equation}
The dependence on the initial time, $t_0$, is a gauge transformation known as the \textit{Floquet gauge}. The Floquet Hamiltonian is a consequence of the observation $U(t_0 + n T, t_0)= U(t_0 + T ,t_0)^n$. A classical version of (\ref{eqn:floquet}) also exists; for vectorized quantum systems, exchange $-i H_F[t_0] \mapsto \mathbf{L}_F[t_0]$ in the manner outlined in Appendix~\ref{apdx:vectorize}.

Take the eigendecomposition of the Floquet operator such that $\mathbf{L}_F[t_0] \boldsymbol{\xi}_j(t_0) = \varepsilon_j \boldsymbol{\xi}_j(t_0)$ for $j=1,2,\dots,N^2-1$. Floquet's theorem is the observation that the $T$-periodic fast-time dynamics can be separated from the slow-time dynamics governed by $\mathbf{L}_F[t_0]$. The theorem asserts that attaching the periodic fast-time propagation to the eigenvectors, $\boldsymbol{\xi}_j(t + T) = \boldsymbol{\xi}_j(t)$, provides a complete set of solutions for the dynamics. In this Floquet basis, the general time-evolution of an initial state $\mathbf{x}(t_0)$ is given by 
\begin{equation} \label{eqn:floquetsoln}
    \mathbf{x}(t) = \sum_{j=1}^{N^2-1} \boldsymbol{\xi}_j(t) e^{\varepsilon_j (t-t_0)} c_j \,,
\end{equation}
where the $\boldsymbol{\xi}(t)$ are known as \textit{Floquet modes}, the $\varepsilon_j$ are known as \textit{quasi-energies}, and the coefficients $c_j$ are the coordinates of $\mathbf{x}(t_0)$.

Comparing (\ref{eqn:eigcontinuous}) with (\ref{eqn:floquetsoln}), it is evident the DMD algorithm (\ref{eqn:dmd}) can provide a framework for the numerical approximation of Floquet modes and quasi-energies using DMD modes and eigenvalues~\cite{Mezic2016}. We refer to this re-interpretation of the DMD algorithm as Floquet~DMD. Suppose a size-$M$ measurement record of the trajectory $\mathbf{x}(t)$ is obtained from a non-autonomous system with a known period $T$. Further suppose that the first $s$ measurements are contained within $[0,T)$, the second within $[T, 2T)$, and so on under the constraint $t_r \equiv t_{ns + r} \mod T$ for $r=1,2,\dots,s$. Consider the reshaped snapshot matrices for this stroboscopically-measured data,
\begin{eqnarray}
    \mathbf{X} =
    \begin{bmatrix}
        \mathbf{x}_1 & \mathbf{x}_{s+1} & \dots & \mathbf{x}_{(\floor{(M-1)/s}-1)s+1} \\
        \mathbf{x}_2 & \mathbf{x}_{s+2} & \dots & \mathbf{x}_{(\floor{(M-1)/s}-1)s+2} \\
        \vdots & \vdots & & \vdots  \\
        \mathbf{x}_s & \mathbf{x}_{2 s} & \dots & \mathbf{x}_{(\floor{(M-1)/s}-1)s+s}
    \end{bmatrix}, \nonumber \\  \label{eqn:snapshotfloquet}
    \mathbf{X'} =
    \begin{bmatrix}
        \mathbf{x}_{s+1} & \mathbf{x}_{2s+1} & \dots & \mathbf{x}_{(\floor{(M-1)/s})s+1} \\
        \mathbf{x}_{s+2} & \mathbf{x}_{2s+2} & \dots & \mathbf{x}_{(\floor{(M-1)/s})s+2} \\
        \vdots & \vdots & & \vdots  \\
        \mathbf{x}_{2s} & \mathbf{x}_{3 s} & \dots & \mathbf{x}_{(\floor{(M-1)/s})s+s}
    \end{bmatrix}
\end{eqnarray}
where $\mathbf{X}, \mathbf{X}' \in \R^{s (N^2-1) \cross \floor{(M-1)/s}}$. In this way, the columns of $\mathbf{X}$ exist in the span of a discretization of the $T$-periodic Floquet modes, $\{\boldsymbol{\xi}_j(t)\}_{j=1}^{N^2-1}$. These Floquet modes evolve with a single phase given by the quasi-energy. From the snapshot matrices, the DMD algorithm constructs an optimal propagator in terms of DMD modes and eigenvalues that provide numerical approximations to the Floquet modes and quasi-energies, respectively.

\subsubsection{Example 2} \label{sec:example2}
%
\begin{figure}
    \includegraphics[width=0.95\textwidth]{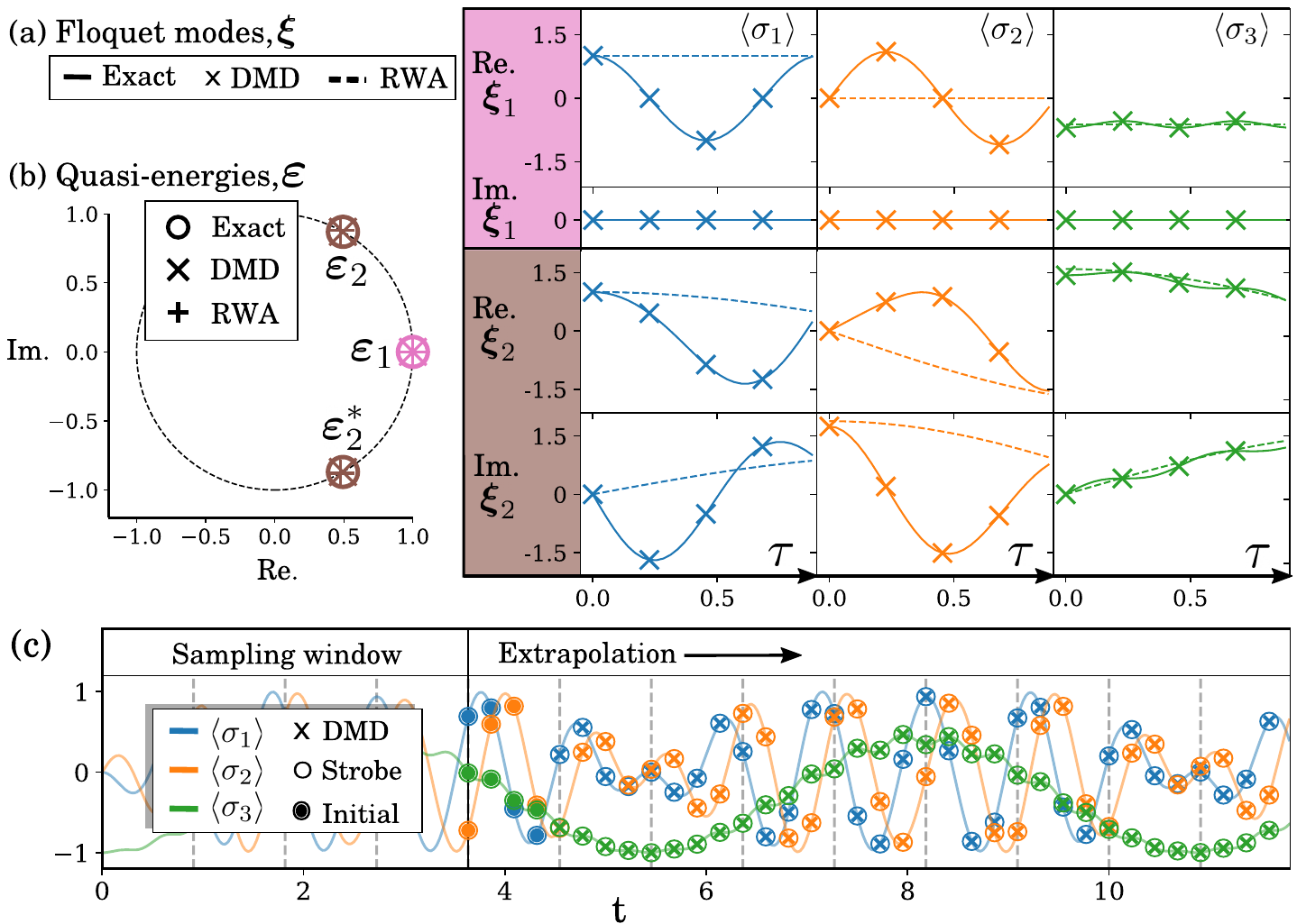}
    \caption{1
        A qubit with intrinsic period $1$ is driven by a slightly off-resonance frequency with period $T = 1/1.1$.  Equi-spaced stroboscopic measurements are sampled at a frequency of $4/T$ during a total sampling window of $4 T$ to construct a model using Floquet DMD. Figures~\ref{fig:floquetmodes}(a-b) compare Floquet modes and quasi-energies to the eigenmodes and eigenvalues from DMD and the rotating wave approximation (RWA). The columns in Figure~\ref{fig:floquetmodes}(a) are indexed by the coordinate of the state $\mathbf{x}(t) = \Tr(\rho \sigma_j)$. The horizontal axes are labelled by $\tau \equiv t \mod T$.  Measurement data from a fifth period is used to provide initial conditions for the model-based extrapolation in Figure~\ref{fig:floquetmodes}(c).
        }
    \label{fig:floquetmodes}
\end{figure}

Consider the two-level quantum system from Section~\ref{sec:example1} with $H(t) = \pi \sigma_3 + u(t) \sigma_1$ driven by a control $u(t) = \cos(2 \pi \omega_D t)$ with $\omega_D = 1.1$. Suppose the system is stroboscopically measured at $4$ times per period over a total $4T$ (such that $M=16$ and $s=4$). Construct the snapshot matrices (\ref{eqn:snapshotfloquet}) from this measurement record. In Figure~\ref{fig:floquetmodes}(a,b), the DMD modes and eigenvalues from applying the Floquet DMD algorithm are compared with the exact Floquet modes, exact quasi-energies, the rotating wave approximation (RWA) modes, and the RWA eigenvalues (see Equation~\ref{eqn:rwa}). Because $\omega_D = 1.1$ is close to resonance, the RWA is valid and able to capture the quasi-energies. However, RWA does so by sacrificing the fast-time dynamics. In contrast, the Floquet DMD resolves both the fast and slow time-scales using the stroboscopic measurement data.  Moreover, Floquet DMD is not limited to a regime around resonance. The Floquet DMD provides a linear reduced order model that approximates the exact linear dynamics of any periodically-driven quantum dynamical system. This implies the method is well-suited to extrapolation, as shown in Figure~\ref{fig:floquetmodes}(c). 

The Floquet DMD inherits all the model-reduction advantages of the DMD algorithm. However, one drawback of Floquet DMD is that control is internal to the constructed model. As such, the solution can assist control strategies involving switched systems, but cannot generalize to unseen control inputs.

\subsection{biDMD and Average Hamiltonian Theory} \label{sec:ffm}
%
\begin{figure}[t]
    \includegraphics[width=\textwidth]{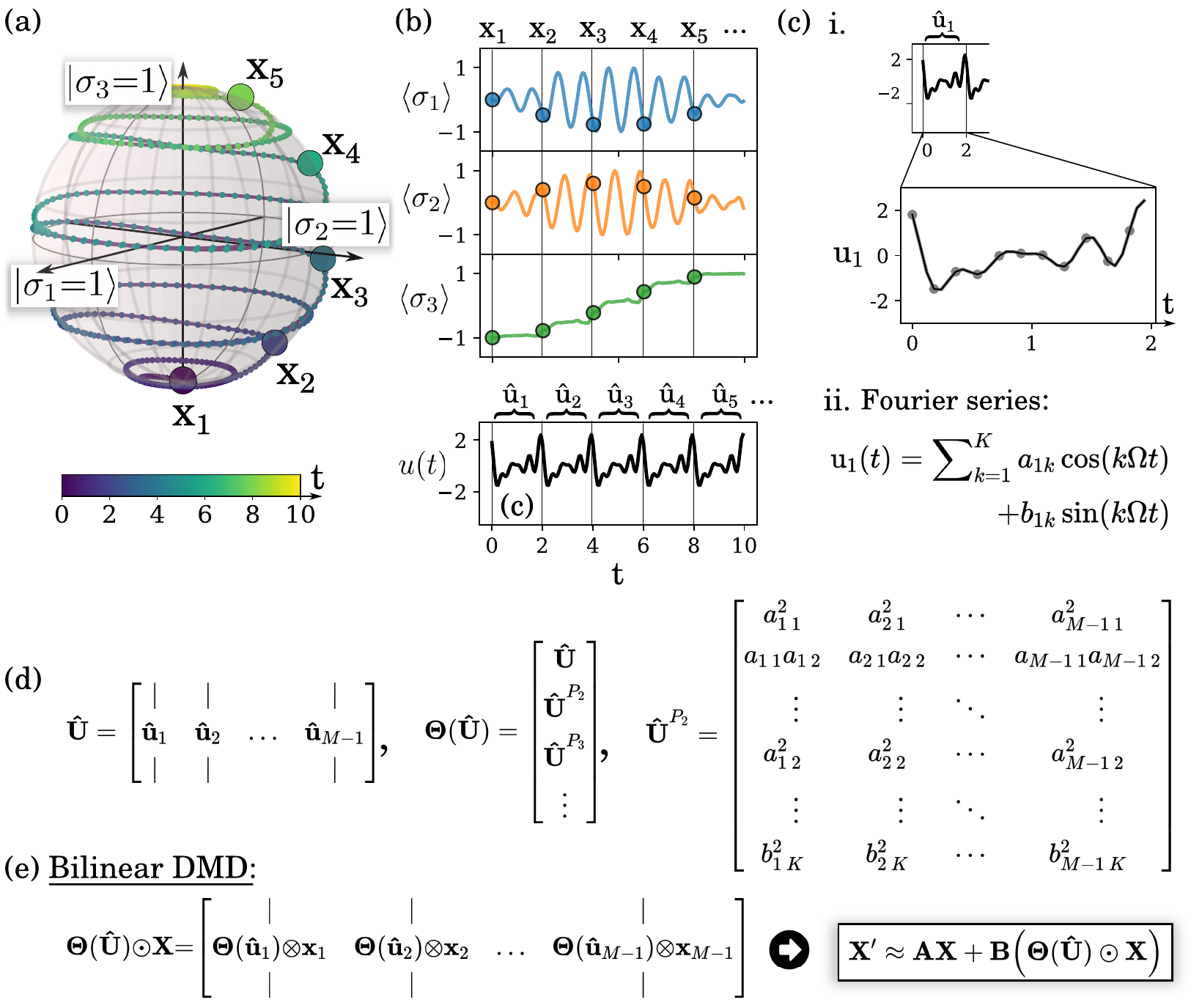}
    \centering
    \caption{
         Stroboscopic measurements of a qubit driven by a linearly-polarized semi-classical drive $\textrm{u}(t)$ (Hamiltonian: $H(t) = \pi \sigma_z + \textrm{u}(t) \sigma_x$) are shown on the Bloch sphere in Figure~\ref{fig:main2}(a). The corresponding Pauli-spin measurements are shown in Figure~\ref{fig:main2}(b). Control is incorporated by computing Fourier coefficients of each intra-stroboscopic control input (Figure~\ref{fig:main2}(c)): $\mathbf{\hat{u}}_n = [a_{n1},a_{n2},\dots,a_{nK},b_{n1},b_{n2},\dots,b_{nK}]\T$. Potentially nonlinear combinations of the control coefficients are assembled into a control data matrix in Figure~\ref{fig:main2}(d). The stroboscopic bilinear Dynamic Mode Decomposition (biDMD, Figure~\ref{fig:main2}(e)) is a regression-based algorithm that uses the assembled data matrices and control input from stroboscopic measurements to learn the intrinsic dynamics,~$\mathbf{A}$ and the control,~$\mathbf{B}$, for the bilinear dynamics.
         }
     \label{fig:main2}
\end{figure}

Average Hamiltonian Theory (AHT)~\cite{Vandersypen2005,Brinkmann2016} is a method with origins in the NMR community that asserts that the dynamics of a quantum dynamical system driven by a time-dependent control can be described by the average effect of the field over a period $T$ of its oscillation. In AHT, a known system Hamiltonian is expanded analytically according to the Magnus series~\cite{Magnus1954,Blanes2009}. In this section, we show how AHT can be combined with the ideas of Floquet DMD from Section~\ref{sec:floquet} to provide a framework for biDMD (\ref{eqn:bidmd2}) in the case of stroboscopic measurement data. This allows us to extend model-free optimal control to the limit of stroboscopic measurements. We refer to this interpretation of biDMD as AHT-biDMD.

Consider a periodic bilinear dynamical system
\begin{equation}
    \frac{\partial}{\partial t} \mathbf{x}(t) = \mathbf{L}(t)\mathbf{x}(t) = \left( \mathbf{L}_0 + u(t) \mathbf{L}_u \right) \mathbf{x}(t)
\end{equation}
such that $\mathbf{L}(t + T) = \mathbf{L}(t)$. Define $\Omega := 2\pi/T$. Recall the existence of the constant Floquet operator $\mathbf{L}_F[t_0]\mathbf{x}(t_0) = \mathbf{x}(t_0 + T)$. The main idea of this section is to approximate $\mathbf{L}_F[t_0]$ using a high-frequency perturbative expansion which will motivate a way to include control as a parameter in the spirit of biDMD. There are a number of perturbative methods to find the Floquet operator in the limit of a high-frequency drive; we will use the familiar Magnus expansion as used in AHT which allows us to represent the constant Floquet operator $\mathbf{L}_F[t_0]$ as a series of constant operators~\cite{Bukov2015},
\begin{equation} \label{eqn:magnus}
    \mathbf{L}_F[t_0] = \sum_{j=0}^{\infty} \mathbf{L}_F^{(j)}[t_0]
\end{equation}
where $L_F^{(n)}[t_0]$ means that the operator appears in the series expansion with a pre-factor $\Omega^{-n}$. By inserting the Fourier series $u(t) = \sum_{k=1}^{K} a_k \cos(k \Omega t) + b_k \sin(k \Omega t)$ into each $\mathbf{L}^{(j)}_F[t_0]$ in (\ref{eqn:magnus}), $j=0,1,\dots, J-1$, we obtain a bilinear model up to terms of order $T^{-J}$ in which controls enter the Floquet operator as polynomials of the constant Fourier series coefficients. An analytic example is shown in Appendix~\ref{apdx:aht}. The expansion presumes these coefficients remain constant across a single period $[(n-1)T, nT]$ but allows for changes between periods. The bilinearization of the propagator through the basis decomposition of the control is motivated by analytic Floquet-Fourier methods~\cite{Shirley1965,Chu2004,Deconinck2006}.

To apply AHT-biDMD, construct the snapshot matrices $\mathbf{X}$ and $\mathbf{X}'$ as in (\ref{eqn:snapshots}) using the size-$M$ measurement record of the trajectory $\mathbf{x}(t)$ where $\mathbf{x}_n$ are measured stroboscopically, $t_n = t_0 + (n-1)T$. Additional intra-stroboscopic measurements can be included using the methods outlined in Section~\ref{sec:floquet}. Next, suppose controls are restricted to $u_n(t) = \sum_{k=1}^{K} a_{nk} \cos(k \Omega t) + b_{nk} \sin(k \Omega t)$ where $n$ indexes the relevant step $[(n-1)T, nT]$ of the applied control. Define a control snapshot in terms of the coefficients, $\mathbf{\hat{u}}_n = [a_{n1},a_{n2},\dots,a_{nK},b_{n1},b_{n2},\dots,b_{nK}]\T$, such that the snapshot matrix is now
\begin{equation}
    \mathbf{\hat{U}} = \begin{bmatrix}
        | & | & & | \\
        \mathbf{\hat{u}}_{1} & \mathbf{\hat{u}}_{2} & \dots & \mathbf{\hat{u}}_{M-1} \\
        | & | & & |
    \end{bmatrix}\,.
\end{equation}
Furthermore, define the polynomial library matrix to be
\begin{equation} \label{eqn:library}
    \boldsymbol{\Theta}(\mathbf{\hat{U}}) = \begin{bmatrix}
        \mathbf{\hat{U}}\\
        \mathbf{\hat{U}}^{P_2}\\
        \mathbf{\hat{U}}^{P_3}\\
        \vdots
    \end{bmatrix}\,, 
\end{equation}
where e.g. $\mathbf{\hat{U}}^{P_2}$ denotes all quadratic nonlinearities of the control coefficients,
\begin{equation}
    \mathbf{\hat{U}}^{P_2} = \begin{bmatrix}
          a_{1\,1}^2   &  a_{2\,1}^2   &   \cdots   &   a_{M-1\,1}^2 \\
          a_{1\,1}a_{1\,2}   &  a_{2\,1}a_{2\,2}   &   \cdots   &   a_{M-1\,1}a_{M-1\,2} \\
          \quad\vdots & \quad\vdots & \ddots  & \quad\vdots \\
          a_{1\,2}^2   &  a_{2\,2}^2   &   \cdots   &   a_{M-1\,2}^2 \\
          \quad\vdots & \quad\vdots & \ddots & \quad\vdots \\
          b_{1\,K}^2   &  b_{2\,K}^2   &   \cdots   &   b_{M-1\,K}^2 
    \end{bmatrix}\,.
\end{equation}
Finally, use the library to construct the bilinear operator as in (\ref{eqn:snapshotbi}), $\boldsymbol{\Theta}(\mathbf{\hat{U}}) \odot \mathbf{X}$. Now, AHT-biDMD is the approximation of the discrete time dynamics by the model
\begin{equation} \label{eqn:bidmdffm}
    \mathbf{X}' \approx \mathbf{A} \mathbf{X} + \mathbf{B} \left( \boldsymbol{\Theta}(\mathbf{\hat{U}}) \odot \mathbf{X} \right) \,.
\end{equation}
The algorithm proceeds as in biDMD in Section~\ref{sec:bidmd}.

\begin{figure}
    \includegraphics[width=\textwidth]{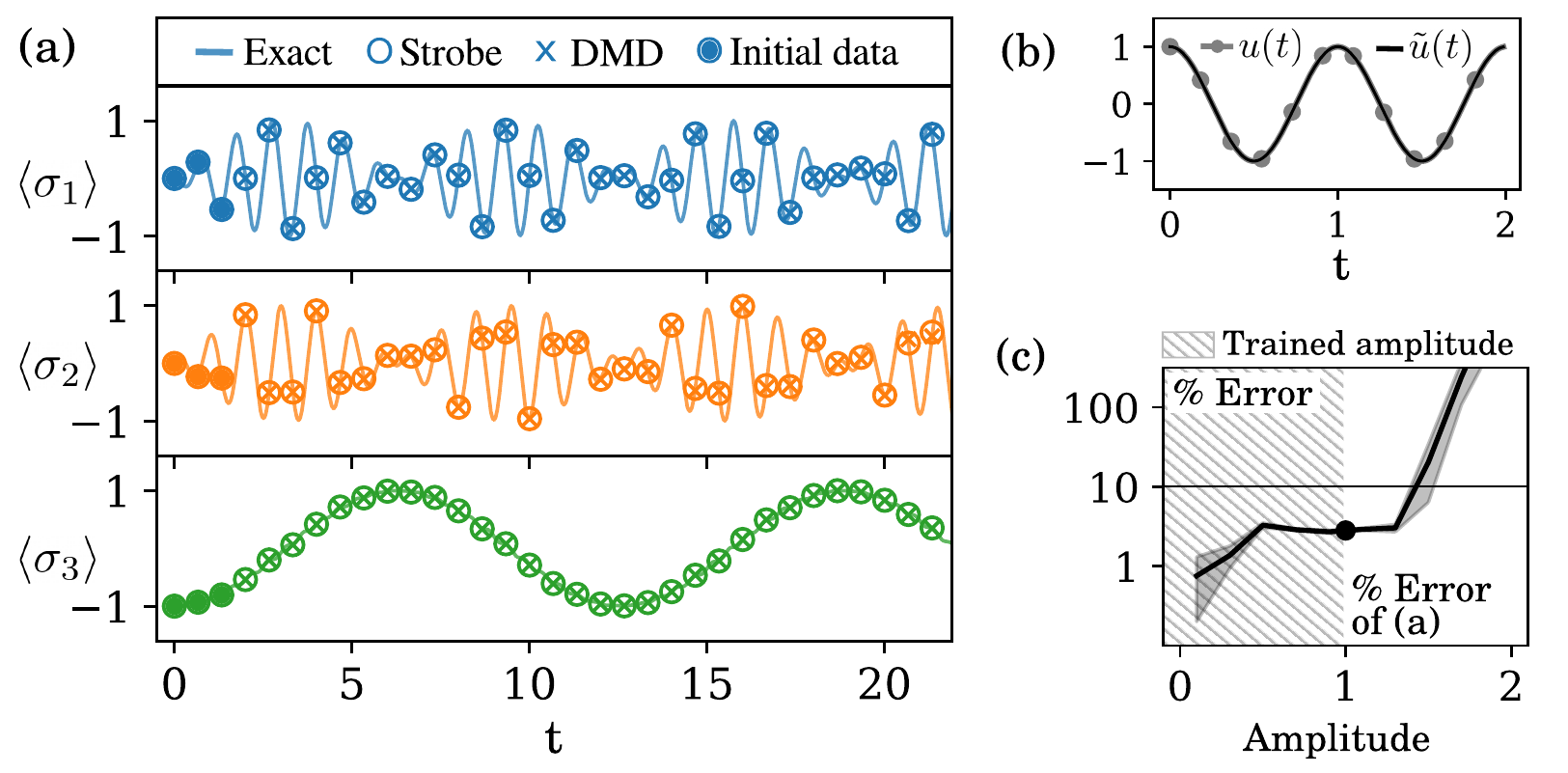}
    \caption{Figure~\ref{fig:res_control}(a) shows the true trajectory compared to the biDMD predictions for a two-level system driven at resonance. Figure~\ref{fig:res_control}(b) plots the control over a single period and shows an exact match with the truncated Fourier series whose coefficients make up the control input of the biDMD model. Figure~\ref{fig:res_control}(c) shows the error of (a) and for additional amplitudes of the same control.
    }
    \label{fig:res_control}
\end{figure}
\begin{figure}
    \includegraphics[width=\textwidth]{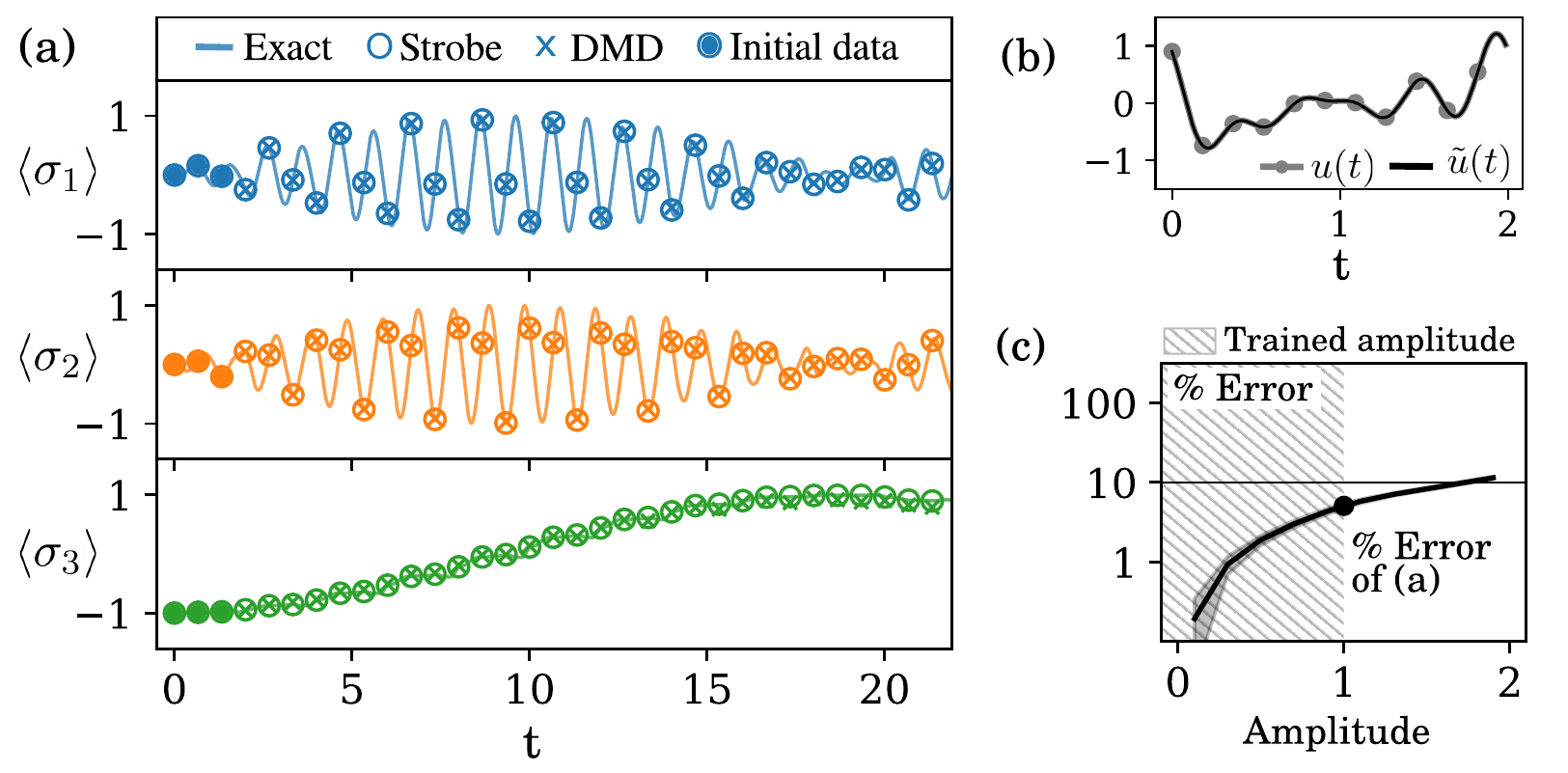}
    \caption{Figure~\ref{fig:known_control}(a) shows the true trajectory compared to the biDMD predictions for a two-level system driven by a random control. Figure~\ref{fig:known_control}(b) plots the control over a single period and shows an exact match with the truncated Fourier series whose coefficients make up the control input of the biDMD model. Figure~\ref{fig:known_control}(c) shows the error of (a) and for additional amplitudes of the same control.
    }
    \label{fig:known_control}
\end{figure}
\subsubsection{Example 3} \label{sec:example3}
%
\begin{figure}
    \includegraphics[width=\textwidth]{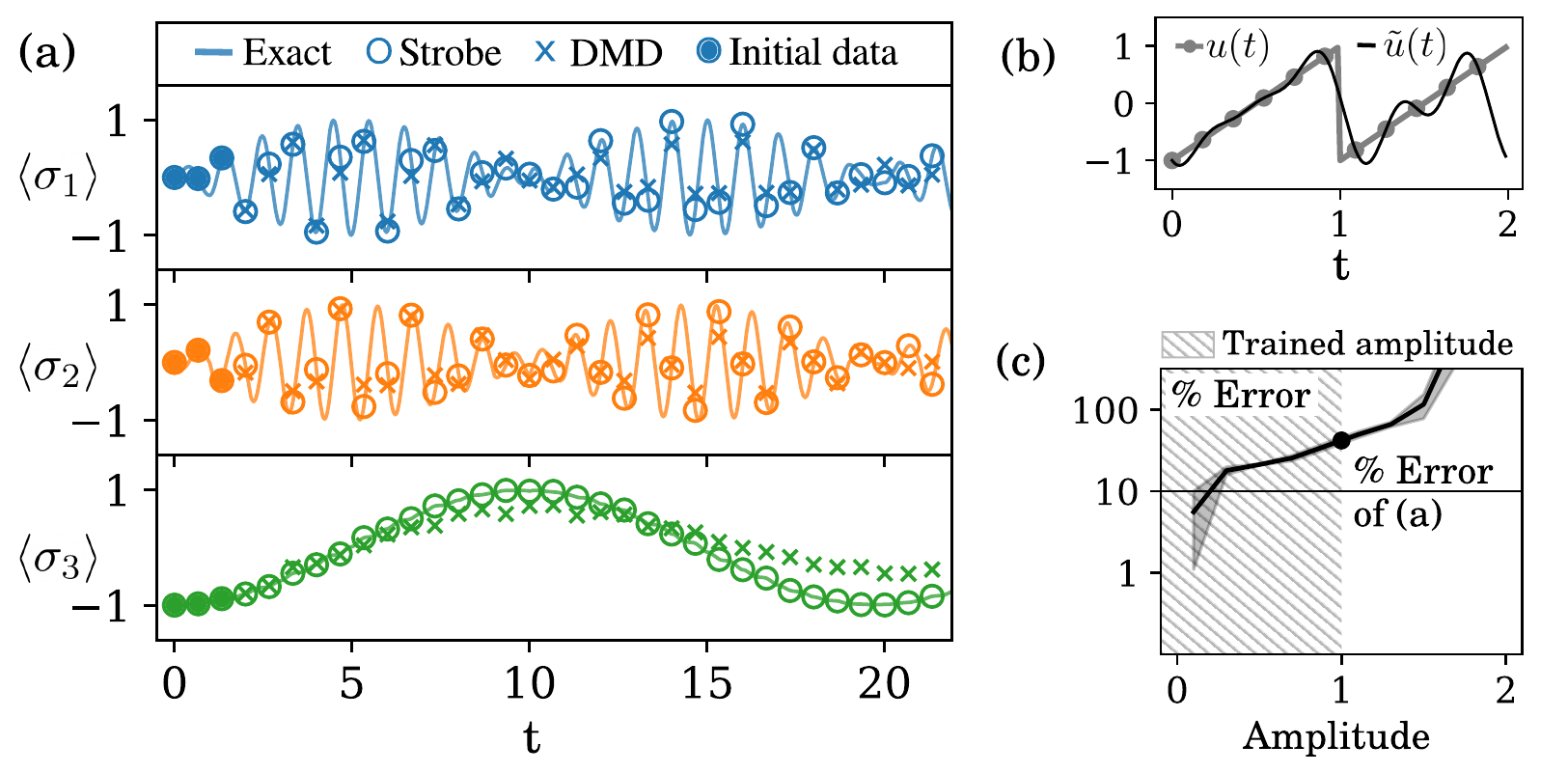}
    \caption{Figure~\ref{fig:saw_control}(a) shows the true trajectory compared to the biDMD predictions for a two-level system driven by a sawtooth control. Figure~\ref{fig:saw_control}(b) plots the control over a single period and compares to the approximation via the truncated Fourier series whose coefficients make up the control input of the biDMD model. Figure~\ref{fig:saw_control}(c) shows the error of (a) and for additional amplitudes of the same control.}
    \label{fig:saw_control}
\end{figure}

Consider again the two-level quantum system with $H(t) = \pi \sigma_3 + u(t) \sigma_1$. Restrict the control to the span
\begin{equation} \label{eqn:controlspan}
    u(t) = \sum_{k=1}^{5} a_{k} \cos(k \Omega t) + b_{k} \sin(k \Omega t), \quad \Omega = \frac{1}{2} \frac{2\pi}{T} = \pi \,.
\end{equation}
Leveraging the linearity of the biDMD model, assemble the measurement trajectory data as a horizontal stack of numerical experiments driven by pure tones. Suppose the trajectories are measured over 5 periods of applied control in which each of $a_{k}$ and $b_{k}$ are taken separately to be $0, 0.1, \dots, 1$. Include additive Gaussian noise with $\sigma=10^{-2}$. From this control data, construct the library snapshot matrix (\ref{eqn:library}) up to second-order nonlinearities. In the trajectories, we also include additional intra-stroboscopic measurements using the methods of Section~\ref{sec:floquet}. From all of these horizontally-stacked measurement trajectories and the polynomially-extended control coefficients, assemble the snapshot matrices according to~(\ref{eqn:bidmdffm}), and apply the biDMD algorithm to obtain a model for the system.

In Figures~\ref{fig:res_control},~\ref{fig:known_control},~and~\ref{fig:saw_control} we look at the predictive capabilities of the biDMD model over $10$ control periods (twice the training time) for certain representative control pulses. Part~(b) of each figure shows the input control plotted alongside the Fourier-series signal (\ref{eqn:controlspan}) that best approximates the applied control. Part~(c) quantifies the percent error of the measured versus predicted trajectories over a range of amplitudes assuming that the input control pulse remains in a fixed shape.

In particular, Figure~\ref{fig:res_control} shows the case of a resonance drive. The model is successful up to a large amplitude limit. This can be understood by recalling that the validity of the Magnus expansion requires a control period that is ``small enough''--a condition that exists in an inverse relationship with the amplitude of the applied drive. Figure~\ref{fig:known_control} shows a successful prediction for the case of an arbitrary multi-frequency control in the span of the truncated Fourier series (for simplicity, we hold the same coefficients across all periods). Finally, Figure~\ref{fig:saw_control} shows a sawtooth drive that is only approximately captured by the truncated Fourier series input in the biDMD model; as a result, the model predicts with a reduced accuracy.

\section{Conclusion} \label{sec:conclusion}
The success of quantum optimal control (QOC) is critically dependent on accurate quantum system identification. In cases where theoretical models are inaccurate or unknown, data-driven system identification is essential for practical QOC. Established ideas from machine learning and regression algorithms like the dynamic mode decomposition have advantages that can be useful to quantum technologies. Like the innovation of the bilinear dynamic mode decomposition (biDMD) in this paper, the ideas must first be adapted to accommodate the underlying mathematics and structure of quantum dynamical systems. Much like its forerunner, the DMDc algorithm~\cite{Proctor2016} for the case of direct actuation in classical control systems, biDMD provides a data-driven and equation-free framework for use with QOC strategies. We have demonstrated the success of the biDMD formulation on a number of quantum systems, showing that it can successfully leverage time-series data alone to enact control.

The use of the DMD framework for quantum systems brings with it a variety of well-established extensions and improvements beyond the naive algorithms introduced in this paper. Studying the ways these improvements fit into the quantum world provides a myriad of future research directions. First, DMD is by construction a method for reduced-order modelling of high-dimensional systems; however, DMD can also accommodate tensor-network representations of high-dimensional data~\cite{Peitz2020}. DMD can also utilize compressive measurements~\cite{Brunton2015,Bai2020}. As such, biDMD should be studied in systems involving multiple qubits under control. Also, time-delays and their connection to random measurements~\cite{Ohliger2013,Yang2020} indicate ways DMD can increase system dimensionality in the case of sparse sampling and provide a path for DMD to capture non-Markovian dynamics~\cite{Dylewsky2020}. Optimization-based DMD algorithms~\cite{Askham2018,Lange2020} can improve characterization of the DMD system by relaxing data-collection strategies and by incorporating additional physical constrains or modelling assumptions. Such strategies can be used to increase the fidelity of DMD-based models in experimental settings. Because control is exogenous, feedback can also be used to increase the DMD fidelity~\cite{Geremia2002}. Ultimately, the goal is to demonstrate experimentally that biDMD provides one path to data-driven and equation-free model-based feedback controllers for engineering future quantum technologies.  Moreover, it provides a clear connection to Koopman theory for classical system identification and control, illustrating how nonlinear classical dynamics are linearized via an infinite-dimensional Hilbert space of observables~\cite{Koopman1931,Koopman1932,Mezic2005}.

\section*{Acknowledgements}
SLB acknowledges funding support from the Army Research Office ({ARO W}911{NF}-19-1-0045). 
SLB and EK acknowledge support from the Army Research Office ({ARO W}911{NF}-17-1-0306).  
SLB, EK, and JNK acknowledge support from the UW Engineering Data Science Institute, NSF HDR award \#1934292.  
JLD and AG acknowledge support from the National Nuclear Security Administration Advanced Simulation and Computing Beyond Moores Law program (Grant No. LLNL-ABS-795437). This work was partially performed under the auspices of the U.S. Department of Energy by Lawrence Livermore National Laboratory under Contract No. DE-AC52-07NA27344.

\medskip
 \begin{spacing}{.8}
 \small{
 \setlength{\bibsep}{5.pt}
 \bibliographystyle{unsrt}
 \bibliography{bibliography/andy}
 }
 \end{spacing}
\normalsize

\appendix
\section{Vectorization of open quantum dynamics} \label{apdx:vectorize}
In quantum dynamics, a time-independent closed quantum system can be described by a Hermitian operator $H_0$ (the Hamiltonian) acting on a unit vector $\ket{\psi(t)}$ (the wavefunction) in a complex Hilbert space $\mathcal{H}$ using the Schr\"{o}dinger equation,
\begin{equation} 
    \frac{\partial}{\partial t} \ket{\psi(t)} = -\frac{i}{\hbar} H_0 \ket{\psi(t)}\,, \quad \ket{\psi(0)} = \ket{\psi_0} \,.
\end{equation}
In this paper we set $\hbar=1$, choose $H_0$ to have zero trace, and assume that $\mathcal{H}=\C^N$ has finite dimension. The control of the system can be realized using control functions, $u_k(t)$, coupled to time-independent interaction Hamiltonians, $H_k$, leading to a bilinear Schr\"{o}dinger equation,
\begin{equation}
    \frac{\partial}{\partial t} \ket{\psi(t)} = -i \left(H_0 + \sum_k u_k(t) H_k \right) \ket{\psi(t)} \,.
\end{equation}

More generally, open quantum dynamics involves mixtures of pure quantum states. The statistics of measurements of this kind of state can be completely described by a density operator, $\rho$, that is any element in the convex set of all non-negative ($\rho \ge 0$) self-adjoint ($\rho^\dagger= \rho$) linear maps on $\mathcal{H}$ with $\Tr \rho = 1$. The pure state density operator is $\rho(t) = \ket{\psi(t)}\bra{\psi(t)}$. The case of Markovian dynamics of $\rho$ can be described by the action of completely-positive trace preserving maps generated by the Gorini–Kossakowski–Sudarshan–Lindblad (GKSL) equation~\cite{Breuer2010},
\begin{equation} \label{eqn:lindblad}
    \frac{\partial}{\partial t}\rho(t) = \Lb(t) \rho(t) = -i[H(t), \rho(t)] + \frac{1}{2} \sum_{j,k=1}^{N^2-1} c_{jk} \left( [D_j, \rho(t) D_k^\dagger] + [D_j \rho(t), D_k^\dagger] \right) \,,
\end{equation}
where $H(t)$ is a trace-zero Hermitian operator (the system Hamiltonian), $\{D_j\}_{j=1}^{N^2-1}$ is an orthonormal set of complex matrices with trace zero, and $C:=(c_{jk})$ is positive semi-definite. For a closed quantum system $C = 0$, and the equation becomes the quantum Liouville equation.

The Lindbladian, $\Lb$, is sometimes called a super-operator because of its linear action on the operators $\rho$ (in the Liouville limit, $\mathcal{L}$ is instead called the Liouville operator). Several ways to \textit{vectorize} the GKSL equation exist.
We will use the physics language of the Bloch vector (also know as the vector of coherence) to describe passing to a vector ODE.  Using the trace-one constraint of a density matrix, write 
\begin{equation}
    \rho(t) = \frac{\1}{N} + \sum_{j=1}^{N^2-1} \Tr(\rho \sigma_j) \sigma_j
\end{equation}
with $\{\sigma_j\}_{j=1}^{N^2-1}$ as a complete and orthonormal basis for traceless Hermitian operators. Take $x_j := \Tr(\rho \sigma_j)$ so that (\ref{eqn:lindblad}) becomes
\begin{equation}
    \frac{\partial}{\partial t} \mathbf{x}(t) = ( L_H + L_D ) \mathbf{x}(t) + \mathbf{c}
\end{equation}
by projection onto the basis $\{\sigma_j\}_{j=1}^{N^2-1}$ such that
\begin{align}
    &(\mathbf{L}_H)_{j,k} = \sum_\ell \Tr(H\, \sigma_\ell) f_{j \ell k} \\
    &(\mathbf{L}_D)_{j,k} = -\frac{1}{4} \sum_{\ell,m,n} c_{mn} 
                    \left( f_{m \ell j} \left(f_{n \ell k} - i g_{n \ell k} \right) 
                    + f_{n \ell j} \left(f_{m \ell k} + i g_{m \ell k}\right) \right) \\
    &(\mathbf{c})_j =   \frac{i}{N} \sum_{m,n} c_{mn} f_{mnj}
\end{align}
where $\left[\sigma_{j}, \sigma_{k}\right] =  i \sum f_{j k \ell} \sigma_{\ell}$ and $\left\{\sigma_{j}, \sigma_{k}\right\} = \frac{2}{N}\delta_{j k} \1 +\sum_{\ell} g_{j k \ell} \sigma_{\ell}$ are the structure constants of the basis~\cite{Altafini2012, Kurniawan2012}. Identifying $H(t) = H_0 + \sum_k u_k(t) H_k$ with $\mathbf{L}_H = \mathbf{L}_0 + \sum_k u_k(t) \mathbf{L}_k$ results in a bilinear GKSL equation for the vectorized density matrix,
\begin{equation}
    \frac{\partial}{\partial t} \mathbf{x}(t) = ( \mathbf{L}_0 + \sum_k u_k(t) \mathbf{L}_k + \mathbf{L}_D ) \mathbf{x}(t) + \mathbf{c}\,.
\end{equation}
Similarly, the bilinear quantum Liouville equation is
\begin{equation}
    \frac{\partial}{\partial t} \mathbf{x}(t) = ( \mathbf{L}_0 + \sum_k u_k(t) \mathbf{L}_k ) \mathbf{x}(t)
\end{equation}

\section{Average Hamiltonian Theory example} \label{apdx:aht}
Recall the two-level quantum dynamical system (\ref{sec:example1}) with $H(t) = \pi \sigma_3 + u(t) \sigma_1$ where $\{\sigma_j\}_{j=1}^{3}$ are the standard Pauli matrices. The corresponding vectorized Liouville equation is attained via the Bloch representation discussed in Appendix~\ref{apdx:vectorize} where $x_j = \Tr(\rho(t) \sigma_j)$ such that
\begin{equation}
    \frac{\partial}{\partial t} \mathbf{x}(t) = \mathbf{L}(t) \mathbf{x}(t) = 2\pi \mathbf{L}_3 \mathbf{x}(t) + 2 u(t) \mathbf{L}_1 \mathbf{x}(t) \,.
\end{equation}
where
\begin{equation}
    \mathbf{L}_1 = \begin{pmatrix} 0 & -1 & 0 \\ 1 & 0 & 0 \\ 0 & 0 & 0 \end{pmatrix}, \quad
    \mathbf{L}_2 = \begin{pmatrix} 0 & 0 & 1 \\ 0 & 0 & 0 \\ -1 & 0 & 0 \end{pmatrix}, \quad
    \mathbf{L}_3 = \begin{pmatrix} 0 & 0 & 0 \\ 0 & 0 & -1 \\ 0 & 1 & 0 \end{pmatrix}
\end{equation}
are the vectorized Pauli matrices. Consider a fixed-frequency drive $u(t) = u \cos(k \Omega t) + v \sin(k \Omega t)$. The goal is to compute the Floquet operator $\mathbf{L}_F[0]\mathbf{x}(n T) = \mathbf{x}((n+1)T)$, with $T = 2\pi/\Omega$, to second order in $\Omega$ using the Magnus expansion. Take $\mathbf{L}_F^{(n)}[t_0]$ to mean that the operator appears in the expansion with a pre-factor $\Omega^{-n}$. The first few terms in the Magnus expansion are:
\begin{align}
    \mathbf{L}_F^{(0)}[t_0] &= \frac{1}{T}\int_{t_0}^{t_0 + T} \Dt \mathbf{L}(t) \nonumber \\
    \mathbf{L}_F^{(1)}[t_0] &= \frac{1}{2! T}\int_{t_0}^{t_0 + T} \Dt_1 \int_{t_0}^{t_1} \Dt_2 [\mathbf{L}(t_1),  \mathbf{L}(t_2)] \nonumber \\
    \mathbf{L}_F^{(2)}[t_0] &= \frac{1}{3! T}\int_{t_0}^{t_0 + T} \Dt_1 \int_{t_0}^{t_1} \Dt_2  \int_{t_0}^{t_2} \Dt_3 [\mathbf{L}(t_1),  [\mathbf{L}(t_2), \mathbf{L}(t_3)] + (1 \leftrightarrow 3) \,.\nonumber \\
\end{align}
The control appears with the same multiplicity as its attached operator in each non-vanishing commutator. Note that:
\begin{align}
    [\mathbf{L}(t_1),  \mathbf{L}(t_2)] &=  4\pi(u(t_2) - u(t_1))[\mathbf{L}_3, \mathbf{L}_1] \nonumber \\
    [\mathbf{L}(t_1),  [\mathbf{L}(t_2), \mathbf{L}(t_3)]] &= 8 \pi^2 (u(t_3) - u(t_2))[\mathbf{L}_3, [\mathbf{L}_3, \mathbf{L}_1]] \nonumber \\
    & \qquad + 8 \pi u(t_1)(u(t_3) - u(t_2))[\mathbf{L}_1, [\mathbf{L}_3, \mathbf{L}_1]] \,.
\end{align}
As such, the desired example expansion is computed to be:
\begin{equation} 
    \mathbf{L}_F[0] = 2 \pi \mathbf{L}_1 + \frac{4 \pi v}{k \Omega} \mathbf{L}_2 
    - \frac{1}{k^2 \Omega^2} \left(2 \pi (u^2 + 3 v^2) \mathbf{L}_1 + 8 \pi^2 u \mathbf{L}_3 \right) + \mathrm{O}(\Omega^{-3}) \,.
\end{equation}

\end{document}